\documentclass[longauth,traditabstract]{aa}

\usepackage{natbib}
\bibpunct{(}{)}{;}{a}{}{,} 

\usepackage{txfonts}
\usepackage[hyphens]{url}
\usepackage{verbatim}
\usepackage{fixltx2e} 

\usepackage{hyperref}
\hypersetup{pdfauthor={Bagmeet Behera},
					pdftitle={H.E.S.S. discovery of VHE gamma-rays from the quasar PKS 1510-089},
					pdfsubject={H.E.S.S. discovery of PKS 1510-089},
		            pdfkeywords={AGN, quasar, H.E.S.S., PKS 1510-089, gamma-rays, EBL, internal absorption}}
		            
\usepackage{graphicx}
\usepackage[small, labelfont=bf]{caption} 
\usepackage{ctable}                                  
\usepackage{color}

\usepackage{subfig,sidecap}

\begin{document}
\sloppy
\bibliographystyle{aa} 

\captionsetup[figure]{labelsep=period}
\captionsetup[table]{labelsep=period}

\title{H.E.S.S. discovery of VHE $\gamma$-rays from the quasar \mbox{PKS\,1510$-$089}}

\titlerunning{Discovery of VHE $\gamma$-rays from the quasar \mbox{PKS\,1510$-$089}}

\authorrunning{H.E.S.S. Collaboration}

\author{H.E.S.S. Collaboration
\and A.~Abramowski \inst{1}
\and F.~Acero \inst{2}
\and F.~Aharonian \inst{3,4,5}
\and A.G.~Akhperjanian \inst{6,5}
\and G.~Anton \inst{7}
\and S.~Balenderan \inst{8}
\and A.~Balzer \inst{9,10}
\and A.~Barnacka \inst{11,12}
\and Y.~Becherini \inst{13,14}
% ...
}

\offprints{B. Behera,\\
\email{Bagmeet.Behera@desy.de}}

\institute{
Universit\"at Hamburg, Institut f\"ur Experimentalphysik, Luruper Chaussee 149, D 22761 Hamburg, Germany \and
Laboratoire Univers et Particules de Montpellier, Universit\'e Montpellier 2, CNRS/IN2P3,  CC 72, Place Eug\`ene Bataillon, F-34095 Montpellier Cedex 5, France \and
Max-Planck-Institut f\"ur Kernphysik, P.O. Box 103980, D 69029 Heidelberg, Germany \and
Dublin Institute for Advanced Studies, 31 Fitzwilliam Place, Dublin 2, Ireland \and
National Academy of Sciences of the Republic of Armenia, Yerevan  \and
Yerevan Physics Institute, 2 Alikhanian Brothers St., 375036 Yerevan, Armenia \and
Universit\"at Erlangen-N\"urnberg, Physikalisches Institut, Erwin-Rommel-Str. 1, D 91058 Erlangen, Germany \and
University of Durham, Department of Physics, South Road, Durham DH1 3LE, U.K. \and
DESY, D-15735 Zeuthen, Germany \and
Institut f\"ur Physik und Astronomie, Universit\"at Potsdam,  Karl-Liebknecht-Strasse 24/25, D 14476 Potsdam, Germany \and
Nicolaus Copernicus Astronomical Center, ul. Bartycka 18, 00-716 Warsaw, Poland \and
CEA Saclay, DSM/Irfu, F-91191 Gif-Sur-Yvette Cedex, France \and
APC, AstroParticule et Cosmologie, Universit\'{e} Paris Diderot, CNRS/IN2P3, CEA/Irfu, Observatoire de Paris, Sorbonne Paris Cit\'{e}, 10, rue Alice Domon et L\'{e}onie Duquet, 75205 Paris Cedex 13, France,  \and
Laboratoire Leprince-Ringuet, Ecole Polytechnique, CNRS/IN2P3, F-91128 Palaiseau, France
% ...
}

\date{Received 21 January 2013/ Accepted 28 April 2013}

\abstract{The quasar \object{\mbox{PKS\,1510$-$089}} ($z=0.361$) was observed with the H.E.S.S. array of imaging atmospheric Cherenkov telescopes during high states in the optical and GeV bands, to search for very high energy (VHE, defined as $E\geq0.1$\,TeV) emission. VHE $\gamma$-rays were detected with a statistical significance of 9.2 standard deviations in $15.8$\,hours of H.E.S.S. data taken during March and April 2009. A VHE integral flux of \mbox{$I(0.15$\,TeV$<E<1.0$\,TeV$)= (1.0\pm0.2_{\mathrm{stat}}\pm0.2_{\mathrm{sys}})\times10^{-11}$\,cm$^{-2}$s$^{-1}$} is measured. The best-fit power law to the VHE data has a photon index of \mbox{$\Gamma=5.4\pm0.7_{\mathrm{stat}}\pm0.3_{\mathrm{sys}}$}. The GeV and optical light curves show pronounced variability during the period of H.E.S.S. observations. However, there is insufficient evidence to claim statistically significant variability in the VHE data. Because of its relatively high redshift, the VHE flux from \mbox{PKS\,1510$-$089} should suffer considerable attenuation in the intergalactic space due to the extragalactic background light (EBL). Hence, the measured $\gamma$-ray spectrum is used to derive upper limits on the opacity due to EBL, which are found to be comparable with the previously derived limits from relatively-nearby BL Lac objects.
Unlike typical VHE-detected blazars where the broadband spectrum is dominated by non-thermal radiation at all wavelengths, the quasar \mbox{PKS\,1510$-$089} has a bright thermal component in the optical to UV frequency band. Among all VHE detected blazars, \mbox{PKS\,1510$-$089} has the most luminous broad line region (BLR). The detection of VHE emission from this quasar indicates a low level of $\gamma-\gamma$ absorption on the internal optical to UV photon field.
}


\maketitle


\section{Introduction}
Blazars are a composite class of active galactic nuclei (AGN), consisting of BL Lacertae-type objects (BL Lacs) and flat-spectrum radio quasars (FSRQs). They are differentiated by the presence (FSRQs) or the absence (BL Lacs) of strong emission lines in their spectra. The broadband spectra of blazars are dominated by non-thermal emission, characterized by rapid variability \citep[see e.g.,][]{WagnerSJ1995} in all frequency regimes, with high and variable polarization in the radio and optical frequency regimes \citep{Aller2003,Mead1990}. 
More than three dozen blazars have been detected in VHE \mbox{$\gamma$-rays}, the overwhelming majority of which belong to the BL Lac class.

\mbox{PKS\,1510$-$089} is a FSRQ at a redshift of \mbox{$z=0.361$} \citep{BurbidgeKinman1966}, with highly polarized radio and optical emission \citep{Stockman1984}. At the milliarcsecond scale individual components in the radio jet show apparent superluminal motion \citep{Homan2001} as high as $46c$ \citep{Jorstad2005} indicating a small inclination angle to the line of sight and high bulk Lorentz factors. Very long baseline interferometry (VLBI) observations of this highly polarized radio jet shows large misalignment between the milliarcsecond and the arcsecond-scale components \citep{Homan2002}. This can be explained by the high bulk Lorentz factor in the jet, which makes a small jet bending appear much larger in the projected orientation seen by an observer.

The broadband spectrum of this source has a synchrotron component that peaks between millimeter and IR wavelengths. \citet{MM1986} report broad emission lines in the spectrum of \mbox{PKS\,1510$-$089} \citep[confirmed by][]{Tadhunter1993}, as well as a clear UV excess (the ``blue bump") on top of the non-thermal continuum. The blue bump is attributed to thermal emission from the accretion disk. The high energy component in the spectrum extends from soft X-rays to GeV $\gamma$-rays. \mbox{PKS\,1510$-$089} has been extensively monitored in X-rays \citep[e.g.,][]{Jorstad2006} and is known to be a bright $\gamma$-ray emitter from the EGRET era \citep{Hartman1999}. \citet{Kataoka2008} have shown that the quasi-simultaneous broadband spectral energy distribution of \mbox{PKS\,1510$-$089} can be well described by an external Compton model with seed photons from a dusty torus. In the soft X-ray band, \emph{Suzaku} data suggest a hardening in the spectrum, which \citet{Kataoka2008} propose could be due to either a small contribution from the synchrotron self-Compton component, or from bulk-Compton scattered radiation. \emph{Fermi}-LAT measurements of the high energy (HE, defined as \mbox{$100$\,MeV$<E<100$\,GeV}) spectrum of \mbox{PKS\,1510$-$089} in different flux states during 2008--2009 were presented in \citet{Abdo2010_PKS1510-089}. The average HE spectrum (derived from the entire data set presented therein) is well described by a log-parabola model, \mbox{$\mathrm{d}N/\mathrm{d}E\propto (E/E_0)^{-\alpha-\beta~\mathrm{\ln}(E/E_0)}$}, with the following best-fit values for the three free parameters -- the spectral-shape parameters \mbox{$\alpha=2.23\pm0.02$} and \mbox{$\beta=0.09\pm0.01$}, and an integral photon flux above $100$\,MeV of \mbox{$(1.12\pm0.03)\times10^{-6}\mathrm{cm}^{-2}\mathrm{s}^{-1}$}, with the reference energy, $E_0$, fixed at $300\,\mathrm{MeV}$.
In the multiwavelength data presented by \citet{Abdo2010_PKS1510-089} no correlation between the flux variations in the HE and X-ray bands is 
found. The authors report a positive correlation between the HE and the optical band. They find evidence for a 13 day lag between the HE and optical \emph{R}-band light curves (with the HE light curve leading). They argue that this behavior can be used to rule out a change in beaming as the main driver for variability in the source. \citet{Abdo2010_PKS1510-089} make an estimate for the mass of the black hole in this source of \mbox{$5.4 \times 10^8\,M_\odot$}, using a model for the accretion disk temperature profile and the measured UV flux. They deduce a maximum isotropic $\gamma$-ray luminosity of \mbox{$\simeq 2\times10^{48}\,\mathrm{erg\,s}^{-1}$}. Because both these estimates are nearly an order of magnitude smaller than the values they obtain for more distant FSRQs, such as \object{\mbox{3C\,454.3}} or \object{\mbox{PKS\,1502+106}}, they argue that \mbox{PKS\,1510$-$089} could be an atypical FSRQ.
At the same time, the high ratio between \mbox{$\gamma$-ray} and synchrotron luminosities is a typical feature of an FSRQ.

The luminous optical--UV photon fields (broad line emission and the blue bump) in FSRQs can cause substantial absorption of VHE photons by electron-positron pair production \citep[see, e.g.,][]{Donea2003,LiuBai2006,PoutanenStern2010}. If the VHE emitting region were to be embedded within the broad-line region (BLR), immersed in the reprocessed accretion-disk emission, the VHE $\gamma$-rays may not escape the system. On the other hand, VHE emission has been reported from two FSRQs, viz. \object{\mbox{3C\,279}} \citep{Albert2008} and \object{\mbox{PKS\,1222$+$216}} \citep{Aleksic2011}.

\mbox{PKS\,1510$-$089} is a good candidate for a VHE emitting FSRQ because it is a bright GeV blazar with a jet that shows highly relativistic bulk motion (hence beaming effects should be strong). \mbox{PKS\,1510$-$089} was observed with H.E.S.S. (High Energy Stereoscopic System) in March--April 2009 to search for VHE emission, when it was reported to be flaring in the HE and optical domains. MWL data from this period have been published elsewhere, e.g., \citet{Marscher2010} and \citet{Abdo2010_PKS1510-089}. The H.E.S.S. data are presented followed by the relevant optical and \emph{Fermi}-LAT data, used for triggering H.E.S.S. observations and in the discussion section.
\section{Observations and analysis results}
\label{ObsNResults}
\subsection{H.E.S.S.}
\label{HESSObsNResults}
Following reports in March 2009 of flaring activity in \mbox{PKS\,1510$-$089} in the HE domain  \citep{D'Ammado2009a,Pucella2009,Vercellone2009} as well as in the optical frequencies recorded with the ATOM telescope (an optical flaring state was also independently reported by \citet{Villata2009} and \citet{Larionov2009}), it was observed with the H.E.S.S. array \citep{Hinton2004,Aharonian2006a} between \mbox{MJD\,54910} and \mbox{MJD\,54923}. Subsequent observations were performed between \mbox{MJD\,54948} and \mbox{MJD\,54950} following more optical flaring and another HE flare that was reported by \citet{Cutini2009}. A total of $15.8$\,hours (corrected for dead time) of data passing quality cuts 
\begin{table}
\centering 
\caption{\label{t7}H.E.S.S. data and analysis results} 
\label{HESSData} 
\resizebox{1.0\columnwidth}{!}{
\begin{tabular}{l c c c c c c r} 
\hline\hline 
Data Set & MJD & $\langle Z \rangle$\tablefootmark{a} &  Live time & On\tablefootmark{b} & Off\tablefootmark{c} & $\gamma$ & S\tablefootmark{d} \\ 
  &   & ($^\circ$) &  (hours)  & (counts) & (counts) & (counts) & [$\sigma$]\\ \hline
All data & 54910 - 54950 & 22 & 15.8  &  823 & 5750 & 248 & 9.2 \\
4 Tel. & 54910 - 54918 & 23 &  6.7 & 447 & 3172 & 159 & 8.2 \\
\hline 
\hline 
\end{tabular}
} 
\tablefoot{The first row gives all data taken in stereoscopic mode (i.e., $\geqslant 2$ telescopes taking data). The second row gives the subset where all 4 telescopes were operational.\\
\tablefoottext{a}{The time-weighted average zenith angle.}\\
\tablefoottext{b}{Signal + background events around the source position.}\\
\tablefoottext{c}{Background events from off-source region.}\\
\tablefoottext{d}{The statistical significance in standard deviations.}\\
}
\end{table}
\begin{figure}[h!bt]
\resizebox{\hsize}{!}{\includegraphics[width=9.4cm,clip=true, trim=0mm 6mm 5mm 10mm]
{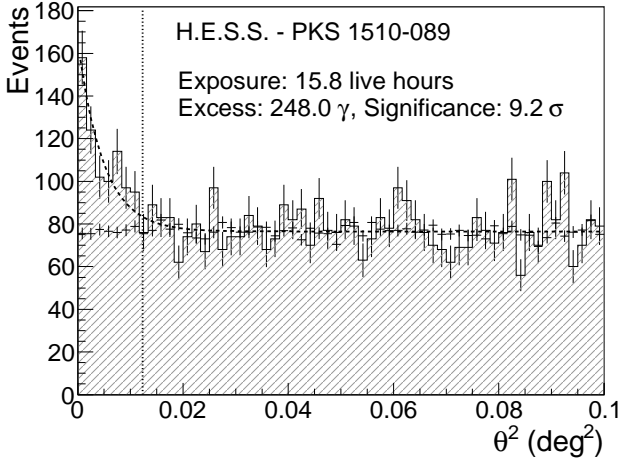}
}
\caption[H.E.S.S. $\theta^2$ distribution of events around the position of \mbox{PKS\,1510$-$089}.]{Distribution of the squared angular distance of $\gamma$-ray candidate events around the position of \mbox{PKS\,1510$-$089}. The angular distribution (in terms of $\theta^2$, the square of the angular distance between the source position and the reconstructed arrival direction) of events around the position of \mbox{PKS\,1510$-$089} is shown. On-source (signal+background) events are shown as hatched histogram, whereas the off-source (background) events are shown as points with error bars. The on-source region is defined by a $\theta^2\leqslant0.0125$\,deg$^2$, shown by the vertical dotted line. The dashed curve is the PSF modeled as a Gaussian, and a constant scaled to the average background level.
}
 \label{ThetaSq_SkyMap}
\end{figure}
\citep{Aharonian2006a} 
were obtained, with zenith angles between \mbox{$14^\circ$ and $41^\circ$}, from all observations.

The data were analyzed using the \emph{Model Analysis} \citep{deNaurois2009}. The relatively high redshift of the source implies that the VHE flux should be strongly attenuated due to the extragalactic background light (EBL) in the optical to IR regime \citep{Nikishov1962}. The amount of EBL extinction increases with the energy of $\gamma$-ray photons, resulting in a steepening of the spectrum measured in the VHE band \citep[see, e.g.,][]{SalmonStecker1998}. Therefore, \emph{loose cuts} \citep{deNaurois2009}, that result in a lower energy threshold, are used in the analysis. A total of $248$ \mbox{$\gamma$-ray} candidates were recorded from the source direction (Table\,\ref{HESSData}, first row), which corresponds to a firm detection with a statistical significance of $9.2\,\sigma$ \citep[following the method of][]{LiMa1983}. The distribution of the squared angular distance of events around the position of \mbox{PKS\,1510$-$089}, determined using the ``reflected background" method \citep{Berge2007}, shows a clear excess in the source region as compared to the background region (Fig.\,\ref{ThetaSq_SkyMap}).
The observed excess is consistent with point-like $\gamma$-ray emission from \mbox{PKS\,1510$-$089}. The best-fit position of the VHE $\gamma$-ray excess is at \mbox{$\mathrm{RA}=15^\mathrm{h}12^\mathrm{m}52\fs2\pm1\fs8_{\mathrm{stat}}\pm1\fs3_{\mathrm{sys}}$} (J2000), \mbox{$\mathrm{Dec.}=-9^\circ6\arcmin21\farcs6\pm26\farcs5_{\mathrm{stat}}\pm20\farcs0_{\mathrm{sys}}$} (J2000). This is compatible with the optical position \citep{Andrei2009} of \mbox{PKS\,1510$-$089}, at a separation of $33\farcs3\pm22\farcs3$, within the statistical and pointing errors of H.E.S.S.
\begin{figure}[!h!bt]
\begin{center}
\subfloat{\resizebox{\hsize}{!}{\includegraphics[clip=true, trim=-6mm 0mm 0mm -3.0mm]{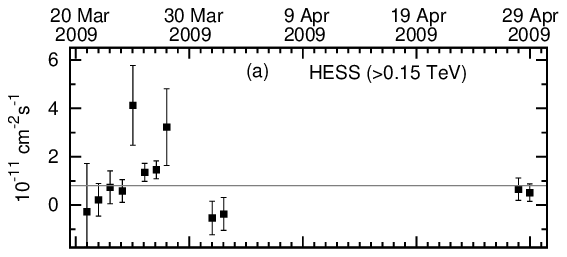}}
 }\\
\vspace{-0.39cm} 
\subfloat{\resizebox{\hsize}{!}{\includegraphics[clip=true, trim=-6mm 0mm 0mm 0.0mm]{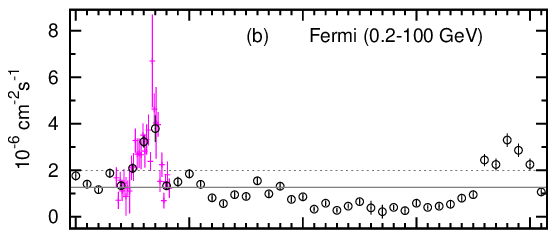}}
}\\
\vspace{-0.39cm}  
\subfloat{\resizebox{\hsize}{!}{\includegraphics[clip=true, trim=-6mm -6.0mm 0mm 0.mm]{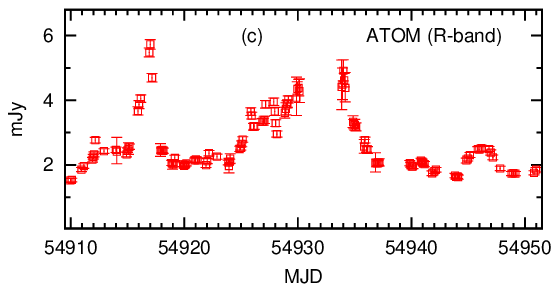}}
}
\caption[Multiwavelength light curves for \mbox{PKS\,1510$-$089} during March--April 2009]{Multiwavelength light curves of \mbox{PKS\,1510$-$089} for the period between MJD\,54910 to MJD\,54951 in terms of integral fluxes in the respective bands along with the $1\sigma$ error bars. Panel (a) shows the VHE light curve (one day bins) from H.E.S.S. The horizontal line is obtained by fitting a constant-source model to the data. Panel (b) displays the HE light curve derived from \emph{Fermi}-LAT data. The black open circles show the integrated fluxes in one day bins, whereas a finer binning of 4 hours is shown in magenta. The solid line is the average level, and the dashed line is the threshold level used for deriving a flare spectrum during the high state around MJD\,54917. In panel (c) the \emph{R}-band optical fluxes measured with ATOM are shown.}
\label{MWL_LC}
\end{center}
\end{figure}

The VHE light curve (in 1 day bins) is shown in the top panel of Fig.\,\ref{MWL_LC}. The best-fit integral flux ($>0.15$\,TeV), obtained by fitting a constant-flux model to the light curve, is  \mbox{$(8.0\pm1.6)\times10^{-12}\mathrm{cm}^{-2}\mathrm{s}^{-1}$} (\mbox{$\chi^2$}$=20.4$\ for $11$ degrees of freedom). The $\chi^2$ test for variability, i.e. for a null hypothesis that the flux is constant, yields a $p\mathrm{-}value\approx0.11$ for this value of the $\chi^2$ test statistic. This is therefore insufficient evidence to claim statistically significant variability (at a $99\%$ confidence level). Because a few bins in the light curve have low statistics (fewer than $10$ counts in the source region and/or background region), the appropriate $p\mathrm{-}value$ was derived from simulations of a large number ($\sim10^5$) of light curves. An average flux of \mbox{$8.0\times10^{-12}\,\mathrm{cm}^{-2}\mathrm{s}^{-1}$} was assumed, and the actual exposure times and instrument effective area in each bin were taken into account. All bins could be thus considered in the test for variability.

\begin{figure}[!h!tb]
\resizebox{\hsize}{!}{\includegraphics[clip=true, trim=-2mm -2.mm 0mm 0mm]
{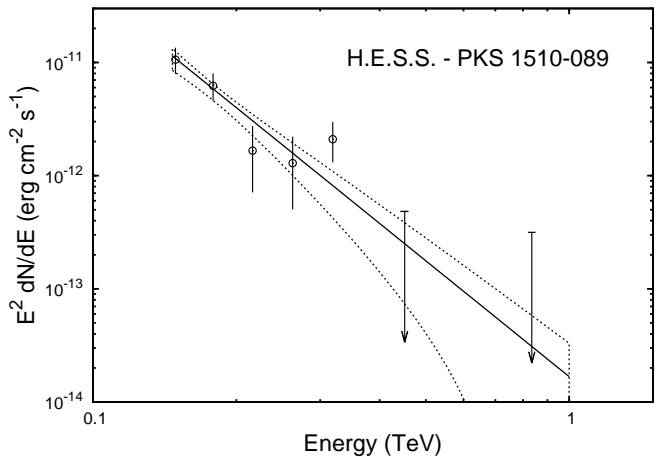}}
\caption[VHE spectrum of \mbox{PKS\,1510$-$089} measured with H.E.S.S. during March 2009]{VHE spectrum of \mbox{PKS\,1510$-$089} measured with the H.E.S.S. instrument, during March 2009. The solid line is the best-fit power law obtained using the forward-folding method (see text for details). The butterfly is the \mbox{$68\%$} confidence band, and the points with errorbars ($1\,\sigma$ statistical errors) are the energy flux. Arrows denote the \mbox{99\%} C.L. upper limits.}
\label{HESS_spectrum}
\end{figure}
To get the best spectral reconstruction, an additional quality criterion was applied, only those data that were taken with the full array of 4 telescopes were accepted (see Table\,\ref{HESSData}, second row). A total of \mbox{$6.7$\,hours} of good quality 4-telescope data were taken during all observations, yielding $159$ \mbox{$\gamma$-ray} candidates from the source direction and a statistical significance of $8.2\,\sigma$. The energy spectrum is derived using a forward-folding technique \citep{Piron2001}.
The analysis threshold, \mbox{$E_\mathrm{thr}\approx0.15$\,TeV}, is given by the energy at which the effective area falls to $10\%$ of its maximum value. For these observations the maximum of the measured differential rate is also at this energy. The likelihood maximization for a power-law hypothesis, \mbox{$\mathrm{d}N/\mathrm{d}E=N_0(E/E_0)^{-\Gamma}$}, in the energy range \mbox{0.15\,TeV}--\mbox{1.0\,TeV}, yields a spectral index of \mbox{$\Gamma=5.4\pm0.7_{\mathrm{stat}}\pm0.3_{\mathrm{sys}}$} and a normalization constant of \mbox{$N_0=(1.1\pm0.2_{\mathrm{stat}}\pm0.2_{\mathrm{sys}})\times10^{-10}\ \mathrm{cm}^{-2}\mathrm{s}^{-1}\mathrm{TeV}^{-1}$} at the decorrelation energy, \mbox{$E_\mathrm{0}=0.18\ \mathrm{TeV}$} (equivalent to a  $\chi^2$ of $10.3$ with $7$ degrees of freedom). The spectral slope is steep compared to other VHE detected blazars, for example, cf. \mbox{$\Gamma=4.11\pm0.68_{\mathrm{stat}}$} for \mbox{3C\,279} \citep{Albert2008}. The reconstructed H.E.S.S. spectrum is shown in Fig.\,\ref{HESS_spectrum}. The integral flux, \mbox{$I(0.15\,\mathrm{TeV}<E<1.0\,\mathrm{TeV})$} \mbox{$=(1.0\pm0.2_{\mathrm{stat}}\pm0.2_{\mathrm{sys}})\times10^{-11}\ \mathrm{cm}^{-2}\mathrm{s}^{-1}$} corresponds to $\approx3\%$ of the Crab Nebula flux in the same band. Fitting a broken-power-law or a power law with an exponential cutoff does not give a better fit. A spectrum obtained from the entire H.E.S.S. data (first row in Table\,\ref{HESSData}) is compatible with the spectrum derived from the high-quality $4$-telescope data, within the statistical errors. The results have been cross-checked using a different analysis method based on multivariate analysis technique \citep[see][and the references therein]{Ohm2009}, and an independent calibration procedure \citep{Aharonian2006a}.
\subsection{Fermi-LAT}
\label{FermiObsNResults}
It is desirable to compare the VHE spectrum to the \emph{Fermi}-LAT \citep{Atwood2009} spectrum during the  contemporaneous period. While the \emph{Fermi}-LAT team has published a spectrum for this source from this period, the integration period used ($\sim1$\,month) is much larger than the shortest flaring timescale seen by Fermi \citep{Abdo2010_PKS1510-089}. The intention here is to obtain a spectrum for the highest state within the \emph{Fermi}-LAT flares (as defined in \citet{Abdo2010_PKS1510-089}) that is contemporaneous with HESS observations. Thus the \emph{Fermi}-LAT data were analyzed using the publicly available \emph{Fermi Science Tools}\footnote{\tiny{\url{http://fermi.gsfc.nasa.gov/ssc/data/analysis/documentation/Cicerone/}}} (v9r23p1-fssc-20111006) and the P7SOURCE\_V6 instrument response functions. The light curve over a contemporaneous period as the H.E.S.S. observations is produced by a binned likelihood analysis retaining photons (the \emph{source} class events) with energies between $200$\,MeV and $100$\,GeV from a region of interest of a radius of $10^\circ$ around the position of \mbox{PKS\,1510$-$089}. All sources from the \emph{Fermi}-LAT two-years point source catalog (2FGL) \citep[][]{2012Nolan2FGL} within an angular distance of $15^\circ$ of \mbox{PKS\,1510$-$089} were modeled simultaneously. \emph{Pass 7 models}\footnote{\tiny{\url{http://fermi.gsfc.nasa.gov/ssc/data/access/lat/BackgroundModels.html}}} of the Galactic and extragalactic backgrounds were used. For these two diffuse backgrounds, the normalizations are treated as free parameters in the likelihood analysis. As mentioned before the \emph{Fermi}-LAT spectrum of this source is best described by a curved log parabola model \citep{Abdo2010_PKS1510-089}. Therefore, a log parabola model is used for the source in the likelihood analysis, using the \emph{gtlike} tool, with the normalization and spectral parameters left free. The light curve derived between \mbox{MJD\,54909.5} and \mbox{MJD\,54951.5} is shown in Fig.\,\ref{MWL_LC}, middle panel.
The average (daily binned) integral flux between $200$\,MeV--$100$\,GeV is \mbox{$(1.26\pm0.03)\times10^{-6}$\,cm$^{-2}$s$^{-1}$}. Two flares are evident, one centered around \mbox{MJD\,54916}, and the second centered around \mbox{MJD\,54948}. Whereas the H.E.S.S. observations during the first HE flare had good coverage and data quality, the H.E.S.S. observations of the second HE flare commenced only after the HE flare had peaked. A finer binning of 4 hours was used around \mbox{MJD\,54916} to precisely trace the development of the flare.

\begin{figure*}[!h!bt]
\centering
\subfloat{\includegraphics[width=8.75cm,clip=true, trim=-2mm -2mm 2mm 0mm]
{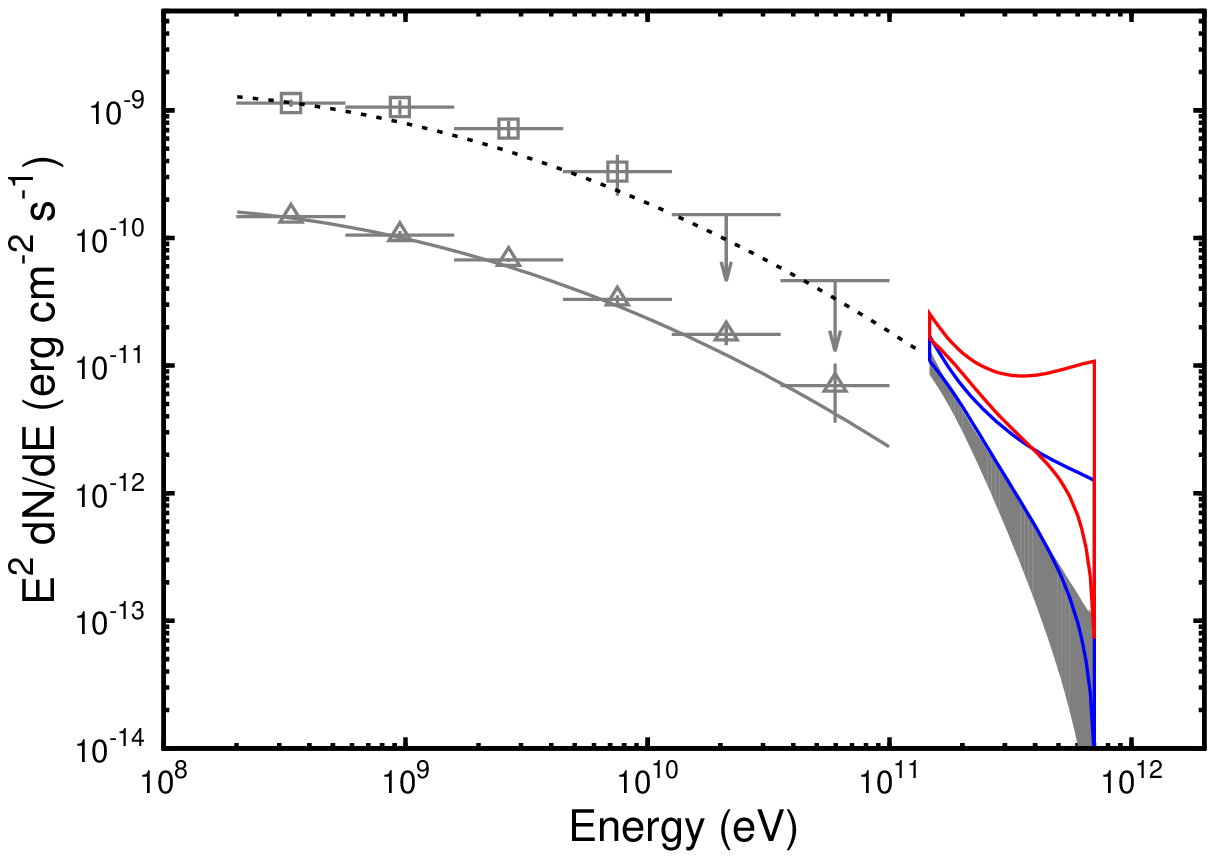}
}
\hspace{0.5cm}
\subfloat{\includegraphics[width=8.75cm,clip=true, trim=0mm -2mm 0mm 0mm]
{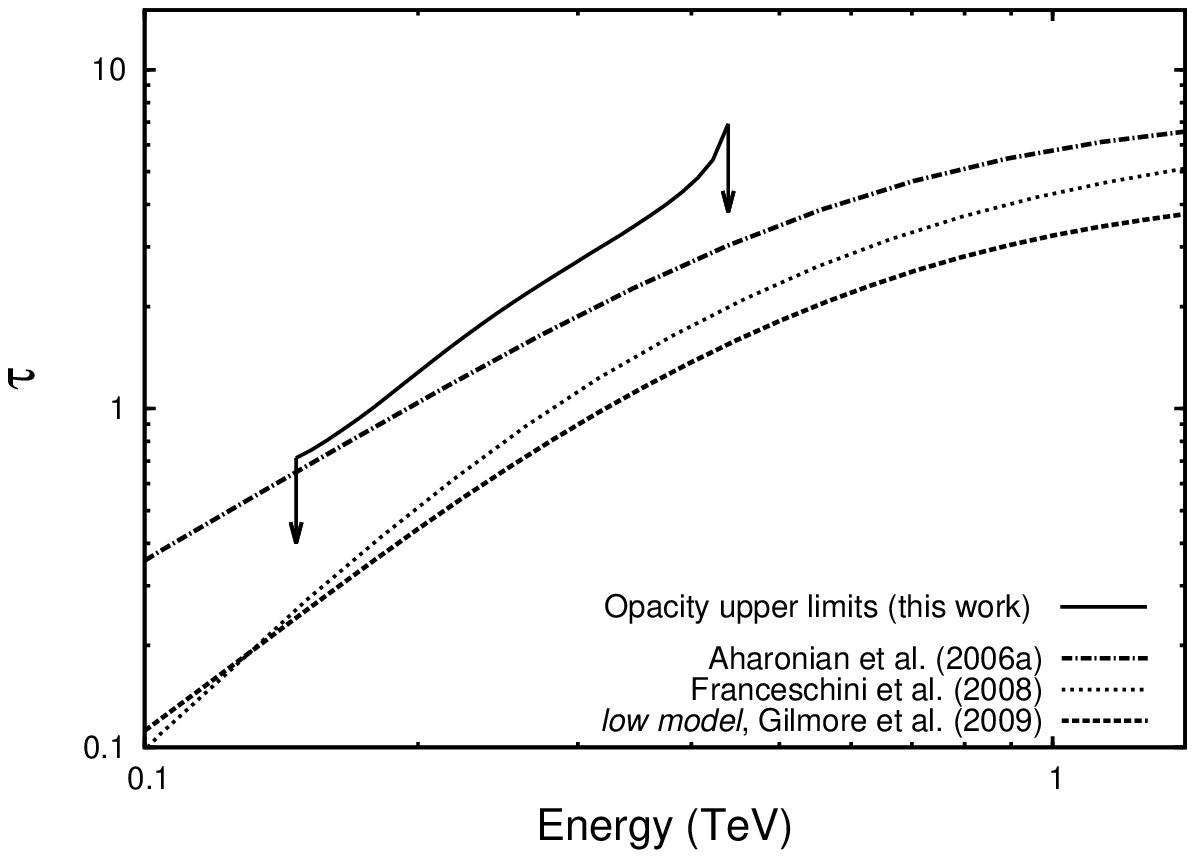}
}
\caption[EBL upper limits derived from the combined \emph{Fermi}-H.E.S.S. $\gamma$-ray spectrum of \mbox{PKS\,1510$-$089}.]{HE and VHE spectra of \mbox{PKS\,1510$-$089} measured with {Fermi}-LAT and H.E.S.S., and upper limits on the EBL induced opacity. \textbf{\emph{Left:}} The filled butterfly shows the H.E.S.S. spectrum as a $1\,\sigma$ confidence band; the blue and red butterflies shows the spectrum corrected for EBL absorption, using the \emph{low model} for the EBL from \citet{Gilmore2009} and the upper-limit EBL model from \citet{Aharonian2006b} respectively. The triangles show the long-term average \emph{Fermi}-LAT spectrum, with error bars denoting $1~\sigma$ statistical errors; the solid curve is the log-parabola model obtained from a binned likelihood analysis of these data (see text for details). 
The open squares show the \emph{Fermi}-LAT spectrum derived from the integration period between \mbox{MJD\,54914.8} and \mbox{MJD\,54917.5} (error bars illustrate $1~\sigma$ statistical errors). Arrows denote upper limits on the flux for those bins where insufficient statistics prohibits a flux measurement. 
The dashed curve is obtained by re-normalizing the long-term average \emph{Fermi}-LAT spectrum to the level of the high-flux state. \textbf{\emph{Right:}} Upper limits on the EBL opacity derived from spectral measurements of \mbox{PKS\,1510$-$089} compared to published models. The solid line with arrows at both ends shows the maximum optical depth, corresponding to upper limits on the EBL. For comparison the opacity corresponding to the previously published H.E.S.S. upper-limit EBL model \citet{Aharonian2006b}, the \citet{Franceschini2008} model and the \emph{low model} from \citet{Gilmore2009} are shown.
}
 \label{FermiExtrapolationHESS}
\end{figure*}

To characterize the range of variations in the HE band, spectra were extracted over two different epochs from \emph{Fermi}-LAT data. A spectrum for the long-term average state was derived from the first two years of \emph{Fermi}-LAT data and a second spectrum was extracted for the high state centered around {MJD\,54916}. The region of interest, the sources modeled together in the binned likelihood analysis, and the parametrization of the source and the background were identical to that used for the light curve generation. The long-term average spectrum was derived from the data taken between \mbox{MJD\,54682}--\mbox{MJD\,55412} (i.e., between 04.08.2008--04.08.2010) for energies above $200\,\mathrm{MeV}$ using a binned likelihood analysis. This results in  a total Test Statistic \citep[TS, e.g., see][]{Mattox1996} of $29440.8$. The likelihood analysis was applied in an iterative way, such that the parameters of the 2FGL sources within $15^\circ$ of \mbox{PKS\,1510$-$089}, and the spectral-shape parameters of \mbox{PKS\,1510$-$089} are fixed in successive steps, leaving only the normalizations of the source and the two diffuse background components free to vary in the last step. The best-fit values for the log-parabola model (\mbox{$\mathrm{d}N/\mathrm{d}E=n_0~[E/E_0]^{-\alpha-\beta~\mathrm{\ln}[E/E_0]}$}) are -- a normalization of $n_0=(1.41\pm0.02)\times10^{-9}~\mathrm{cm}^{-2}\mathrm{s}^{-1}\mathrm{MeV}^{-1}$, and slope parameters of $\alpha=2.21\pm0.03$ and $\beta=0.083\pm0.011$. The choice of the reference energy, $E_0$, does not affect the spectral shape and was thus fixed at $260~\mathrm{MeV}$. These parameters are consistent with the average spectrum presented in \cite{Abdo2010_PKS1510-089}. A spectrum in logarithm bins in energy (using a log-parabola source model) is also derived for the same period. The average and the energy-binned spectrum are shown in Fig.\,\ref{FermiExtrapolationHESS}, left panel.
A \emph{Fermi}-LAT spectrum that is strictly simultaneous to the H.E.S.S. spectrum (that is derived from a total exposure of $6.7$ hours, with gaps in the observations, as well as varying live times for individual exposures) cannot be constructed, because an integration period of at least a few days of \emph{Fermi} observations is required to derive a meaningful spectrum. Moreover, the H.E.S.S. data used for spectral analysis were spread over a period of 8 days, \mbox{MJD\,54910}--\mbox{MJD\,54918}, during which large flux variations were seen at the \emph{Fermi}-LAT energies. Because nearly half of the H.E.S.S. exposure was during the brightest phase of the GeV-flare centered around MJD\,54916, the \emph{Fermi}-LAT data during this period were used to derive a spectrum representing the brightest HE flux state, contemporaneous to the H.E.S.S. observations. Data taken between \mbox{MJD\,54914.8} to \mbox{MJD\,54917.5} were used, where the flux points in the \emph{Fermi}-LAT light curve (in $4$-hour bins) were above a threshold flux of \mbox{$2.0\times10^{-6}\,\mathrm{cm}^{-2}\mathrm{s}^{-1}$} (see Fig.\,\ref{MWL_LC}). This threshold flux was chosen to ensure adequate statistics to construct spectral bins reaching at least \mbox{$10$\,GeV}. The analysis of this data-set yielded a total \mbox{TS$=2354.0$}, with the best-fit values for a log-parabola model given by a normalization of $n_0=(10.5\pm0.7)\times10^{-9}~\mathrm{cm}^{-2}\mathrm{s}^{-1}\mathrm{MeV}^{-1}$, and slope parameters of $\alpha=1.81\pm0.13$ and $\beta=0.161\pm0.048$ (reference energy fixed at $260~\mathrm{MeV}$). The corresponding energy-binned spectrum is shown in Fig.\,\ref{FermiExtrapolationHESS} (left panel). This is roughly an order of magnitude brighter than the average flux, though it should be noted that the spectrum is biased towards the low-energy bins because of the lack of statistics at higher energies. 
\subsection{ATOM}
\label{ATOMObsNResults}
The ATOM telescope (Automatic Telescope for Optical Monitoring), \citet{Hauser2004}, is a $1\,$m class optical telescope, located at the H.E.S.S. site. 
\mbox{PKS\,1510$-$089} is regularly observed with ATOM. From the onset of the H.E.S.S. observation campaigns, the frequency of optical observations was increased to four observations in the \emph{R}-band ($\sim640$\,nm) per night. The \emph{R}-band light curve, with $500$\,s exposure per point and not corrected for galactic extinction, is shown in Fig.\,\ref{MWL_LC}. The optical light curve shows clear and pronounced variability (a reduced $\chi^2=55.3$, with $117$ degrees of freedom from fitting a constant-source model, and the highest flux deviating more than $20\,\sigma$ away from the period average). A prominent flare centered around \mbox{MJD\,54917} was observed, with another flare occurring after \mbox{MJD\,54925} that could not be followed up with H.E.S.S. observations because it happened around a full-Moon phase when H.E.S.S. does not operate. A third brightening was seen around \mbox{MJD\,54946}.

These observations show \mbox{PKS\,1510$-$089} in a relatively high state compared to its average optical brightness, measured with ATOM, between May 2007 and August 2009. For example, in the \emph{R}-band, compared to the long-term average flux of {\mbox{$1.862\pm0.004$\,mJy}, the average during the H.E.S.S. observations was \mbox{$2.597\pm0.009$\,mJy} and a maximum measured flux of \mbox{$5.724\pm0.150$\,mJy}.
\section{Discussion}
\label{Discussion}
The VHE \mbox{$\gamma$-ray} flux of distant ($z\gtrsim0.1$) blazars suffer significant extinction due to pair-production on the diffuse UV-IR photon field in the intergalactic medium \citep{Nikishov1962,Gould1966}. This phenomenon can be used to derive upper limits on the photon density of the UV-IR part of the EBL \citep[e.g.,][]{Aharonian2006b,Mazin2007}.

Here upper limits are derived by comparing the VHE spectra measured with  \mbox{H.E.S.S.} to an extrapolation of the contemporaneous HE spectral measurements obtained with \emph{Fermi}-LAT, following the procedures in, e.g.,  \citet{Georganopoulos2010}, \citet{Aleksic2011} and \citet{Meyer2012}. Modifying \mbox{Equation\,2} from \citet{Aleksic2011}, to include the effect of systematic errors, the \mbox{$95\%$} confidence upper limit on the EBL, expressed in terms of the optical depth, $\tau_{\mathrm{max}}$, is given by
\begin{equation}
\label{Eq1}
\tau_{\mathrm{max}}(E) = \ln \left[\frac{F_{\mathrm{int}}(E)}{(1-\varepsilon)\times(F_{\mathrm{obs}}(E) - 1.64 \times \Delta F_{\mathrm{obs}}(E))} \right],
\end{equation}
where $F_{\mathrm{obs}}(E)$ is the measured flux, $\Delta F_{\mathrm{obs}}(E)$ is the corresponding $1\,\sigma$ statistical error, $F_{\mathrm{int}}(E)$ is the assumed unattenuated intrinsic flux and $\varepsilon$ is the systematic error expressed as a fraction of the measured flux. For a steep spectrum this systematic error (i.e. the uncertainty in the absolute normalization of the flux) is $\sim 25\%$ (i.e. $\varepsilon=0.25$). The systematic error is conservatively considered as a net bias in the flux measurements. It is the factor by which $F_{\mathrm{obs}}(E)$ could overestimate the true flux. The estimation of $\tau_\mathrm{max}$ in \mbox{Equation\,\ref{Eq1}} therefore involves a correction factor of \mbox{$(1-\varepsilon)$}. The extrapolation of the HE spectrum measured with \emph{Fermi}-LAT into the VHE regime is considered as an estimate for the unattenuated spectrum. The large variations in the HE light curve during the H.E.S.S. observing period require the need of contemporaneous spectral measurements. As already mentioned, a strictly simultaneous \emph{Fermi}-LAT spectrum corresponding only to the short exposure of H.E.S.S. (column 2 of Table\,\ref{HESSData}) cannot be obtained because of low statistics in the \emph{Fermi}-LAT data. Even the spectrum obtained during the high flux state (integrated over 2.7d, centered around \mbox{MJD\,54916}) does not provide sufficient statistics at the high-energy end. Therefore, a conservative estimate resulting in the highest possible upper limit on the opacity is derived by re-normalizing the long-term average \emph{Fermi}-LAT spectrum to the level of the high-flux state (dashed curve in Fig.\ref{FermiExtrapolationHESS}, left panel). This is justified because the long-term spectrum gives the best available description of the HE spectral shape up to the highest GeV energies, and changes in spectral shape during flares are small\footnote{\citet{Abdo2010_PKS1510-089} find a harder when brighter trend in the \emph{Fermi}-LAT spectrum when the flux above $0.2\,$GeV is \mbox{$\gtrsim2.4\times10^{-7}$cm$^{-2}$s$^{-1}$}. However, the change in the photon index is small, $\lesssim0.2$.}. This scaled-up spectrum is extrapolated to the VHE regime to estimate the unattenuated flux, $F_{\mathrm{int}}(E)$, used in \mbox{Equation\,\ref{Eq1}}. It is assumed that the spectrum smoothly extends from the HE to the VHE band, without spectral breaks or cutoffs. Any intrinsic spectral breaks resulting in lower fluxes at higher energies would yield a lower opacity from EBL attenuation. These assumptions along with the above mentioned considerations (the inclusion of systematic errors on the H.E.S.S. flux and the selection of the intrinsic spectral shape consistent with the brightest HE flux state) give firm upper limits to the EBL opacity. 
Other sources of opacity, e.g., absorption due to internal photon fields from the BLR or the dusty torus  \citep{WagnerS1995}, would decrease the contribution from EBL extinction to the overall opacity and thus the previous statement still holds. The EBL upper limits derived using \mbox{Equation\,\ref{Eq1}} are shown in \mbox{Fig. \ref{FermiExtrapolationHESS}}, right panel. The EBL limits from this distant FSRQ and the limits in \citet{Aharonian2006b}, that are derived from \mbox{BL Lac} type objects that are relatively nearby, are at a comparable level. 

Owing to their higher bolometric luminosities FSRQs are detected up to larger distances than BL Lacs. Furthermore, due to the bright emission lines in the spectrum of these objects, which the BL Lacs lack by definition, the redshifts of FSRQs can be accurately measured. Thus the potential of putting strong constraints on the EBL extinction is very promising with FSRQs.
However, in the case of \mbox{PKS\,1510$-$089}, other factors currently offset this potential. The H.E.S.S. spectrum turns out to be steep as expected for a high redshift source. This results in a slightly higher systematic error of $25\%$ on the VHE flux measurements (cf. $\sim20\%$ for a Crab-nebula-like spectrum, index $\sim2.4$, \citet{Aharonian2006a}). Furthermore, this sharply falling spectrum makes it difficult to collect good statistics at energies \mbox{$>0.5$\,TeV} and hence the statistical errors in this energy range are also high. The combined effect of the large statistical and systematic errors in the VHE spectrum (denoted by $\Delta F_{\mathrm{obs}}$ and $\varepsilon$ respectively in Equation\,\ref{Eq1}), because of the steep VHE spectral slope of this source, limits the effectiveness of the method used above. Thus only weak EBL limits could be derived.
Another aspect that limits the ability to put strong EBL constraints is the uncertainty in the intrinsic spectral shape.
For the observations presented here there is insufficient evidence for claiming variability in the VHE band. However, because the unabsorbed HE band is highly variable (time scale of hours) it is difficult to obtain an accurate description of the intrinsic spectral shape ($F_{\mathrm{int}}$ in Equation\,\ref{Eq1}) due to the typically multi-day integration time required, hence making it difficult to put stricter EBL constraints. For example, if the estimated $F_{\mathrm{int}}$ at the lowest energies were estimated to be $20\%$ less than the value used, which is perfectly plausible when comparing to the factor of $\sim4$ variations in the \emph{Fermi} fluxes seen during this period, the EBL UL from this source calculated as above would have been more constraining than the \citet{Aharonian2006b} limits.
With more multi-wavelength monitoring of bright FSRQs, such as \mbox{PKS\,1510$-$089} and other FSRQs that have less steep VHE spectrum, it could be possible to obtain a HE and VHE spectrum when the fluxes in neither band varies. This of course is better done with more sensitive Cherenkov telescopes, such as the H.E.S.S. II telescope array, which should provide richer statistics due to their higher sensitivities and larger energy coverage. This should allow putting stronger constraints on the EBL extinction by comparing their quiescent-state HE and VHE spectrum.

\bigskip The luminous broad line emission in the optical-UV band as well as the thermal UV excess in \mbox{PKS\,1510$-$089} indicate an intense internal photon field. This can in principle cause substantial absorption of VHE $\gamma$-rays within the BLR radius, or due to the reprocessed disk emission from the dusty torus at larger distances from the accretion disk. The H.E.S.S. detection of this source implies a low optical depth in the VHE emitting region. This can be tentatively explained by hypothesizing that the VHE emitting region is in an optically thin part of the jet, presumably far outside the BLR, see e.g., \citet[][]{Tanaka2011} and \citet{Tavecchio2011} for discussion on the FSRQ \mbox{PKS\,1222$+$216}. It should be noted that in the study made in \citet{PoutanenStern2010}, where a search is made for signature features of BLR absorption in the HE spectrum of several of the brightest FSRQs detected by Fermi, the HE spectrum of \mbox{PKS\,1510$-$089} is consistent with a negligible amount of absorption due to line emission in BLR clouds. This is consistent with the hypothesis that the $\gamma$-ray emitting zone is in an optically thin region. A detailed discussion of the internal opacity is beyond the scope of this work.

\section{Summary}
\label{Summary}
VHE emission was detected from the flat-spectrum radio quasar \mbox{PKS\,1510$-$089} with a statistical significance of $9.2$\,standard deviations in 15.8 hours of H.E.S.S. data taken during March and April 2009. An integral flux, in the energy regime between \mbox{$0.15$--$1.0\,$\,TeV}, of \mbox{$(1.0\pm0.2_{\mathrm{stat}}\pm0.2_{\mathrm{sys}})\times10^{-11}\ \mathrm{cm}^{-2}\mathrm{s}^{-1}$} was measured, which is $\approx3\%$ of the Crab Nebula flux. The spectrum is extremely steep with a photon index of  \mbox{$\Gamma=5.4\pm0.7_{\mathrm{stat}}\pm0.3_{\mathrm{sys}}$} for a power-law hypothesis.

There is insufficient evidence to claim significant variability in the H.E.S.S. data. However, the multi-frequency data on \mbox{PKS\,1510$-$089} during the H.E.S.S. observation period shows clear and pronounced variability in the HE and optical bands. Using both the \emph{Fermi}-LAT and H.E.S.S. $\gamma$-ray spectrum, from $200\,$MeV to $1\,$TeV, upper limits on the optical depth due to the EBL are derived. The EBL-limits in this work are comparable with the limits in \citet{Aharonian2006b}, that were derived from BL Lac objects that are relatively nearby. FSRQs, such as \mbox{PKS\,1510$-$089}, due to their higher luminosities compared to BL Lacs, can in principle, allow us to probe the EBL density to relatively higher redshifts. With the upcoming H.E.S.S. II telescope array more precise measurements of the spectra of distant blazars are expected, which will allow us to put stronger constraints on the EBL.
\subsubsection*{Acknowledgments}
\tiny 
The support of the Namibian authorities and of the University of Namibia
in facilitating the construction and operation of H.E.S.S. is gratefully
acknowledged, as is the support by the German Ministry for Education and
Research (BMBF), the Max Planck Society, the German Research Foundation (DFG), 
the French Ministry for Research,
the CNRS-IN2P3 and the Astroparticle Interdisciplinary Programme of the
CNRS, the U.K. Science and Technology Facilities Council (STFC),
the IPNP of the Charles University, the Czech Science Foundation, the Polish 
Ministry of Science and  Higher Education, the South African Department of
Science and Technology and National Research Foundation, and by the
University of Namibia. We appreciate the excellent work of the technical
support staff in Berlin, Durham, Hamburg, Heidelberg, Palaiseau, Paris,
Saclay, and in Namibia in the construction and operation of the
equipment. \\
We acknowledge the information provided by the \emph{Fermi}-LAT collaboration
on the \emph{Fermi}-LAT measured fluxes of \mbox{PKS\,1510$-$089} during the $2009$ flaring episodes, which helped
scheduling of the H.E.S.S. observations presented here.\\
This work has been partially supported by the International
Max Planck Research School (IMPRS) for Astronomy \& Cosmic Physics at the University
of Heidelberg. We acknowledge the anonymous referee for providing useful suggestions that improved the clarity.\\ 
\normalsize


\begin{thebibliography}{}

\bibitem[Abdo {et~al.}(2009)]{Abdo2009_LBAS}
Abdo, A.~A., Ackermann, M., Ajello, M., {et~al.} (Fermi Collaboration) 2009, ApJ, 700, 597

\bibitem[{Abdo {et~al.}(2010a)}]{Abdo2010_PKS1510-089}
Abdo, A.~A., Ackermann, M., Agudo, I., {et~al.} (Fermi Collaboration) 2010a, ApJ, 721,  1425

\bibitem[Abdo {et~al.}(2010b)]{Abdo2010_FirstSrcCatalog}
Abdo, A.~A., Ackermann, M., Ajello, M., {et~al.} (Fermi Collaboration) 2010b, ApJS, 188,  405

\bibitem[Aharonian {et~al.}(2006a)]{Aharonian2006b}
Aharonian, F., Akhperjanian, A.~G., Bazer-Bachi, A.~R. {et~al.} (H.E.S.S. Collaboration) 2006a, Nature, 440, 1018

\bibitem[Aharonian {et~al.}(2006b)]{Aharonian2006a}
Aharonian, F., Akhperjanian, A.~G., Bazer-Bachi, A.~R. {et~al.} (H.E.S.S. Collaboration) 2006b, A\&A, 457, 899

\bibitem[Albert {et~al.}(2008)]{Albert2008}
Albert, J., Aliu, E., Anderhub, H. {et~al.} (MAGIC Collaboration) 2008, Science, 320, 1752

\bibitem[Aleksi\'{c} {et~al.}(2011)]{Aleksic2011}
Aleksi\'{c}, J., Antonelli L.~A., Antoranz, P.  {et~al.} (MAGIC Collaboration) 2011, ApJ, 730, L8

\bibitem[Aller {et~al.}(2003)]{Aller2003}
Aller, M.~F., Aller, H.~D. \& Hughes, P.~A. 2003, ApJ, 586, 33

\bibitem[Andrei {et~al.}(2009)]{Andrei2009}
Andrei, A.~H., Souchay, J., Zacharias, N., {et~al.} 2009, A\& A, 505, 385

\bibitem[Atwood {et~al.}(2009)]{Atwood2009}
Atwood, W.~B., Abdo, A.~A., Ackermann, M., {et~al.} 2009, ApJ, 697, 1071

\bibitem[Berge {et~al.}(2007)]{Berge2007}
Berge, D., Funk, S., Hinton, J., {et~al.} 2007, A\& A, 466, 1219

\bibitem[Burbidge \& Kinman(1966)]{BurbidgeKinman1966}
Burbidge, E.~M. \& Kinman, T.~D. 1966, ApJ, 145, 654

\bibitem[{{Cutini} {et~al.}(2009)}]{Cutini2009} Cutini, S., Hays, E. \& {the Fermi Large Area Telescope Collaboration} 2009,  ATel, 2033

\bibitem[D'Ammando {et~al.}(2009)]{D'Ammado2009a}
D'Ammando, F., Bulgarelli, A., Vercellone, S., {et~al.} 2009, ATel, 1957

\bibitem[de~Naurois \& Rolland(2009)]{deNaurois2009}
de~Naurois, M. \& Rolland, L. 2009, Astroparticle Physics, 32, 231

\bibitem[Donea \& Protheroe(2003)]{Donea2003}
Donea, A.-C. \& Protheroe, R.~J. 2003, Astropart. Phys., 18, 377

\bibitem[Franceschini {et~al.}(2008)]{Franceschini2008}
Franceschini, A., Rodighiero, G. \& Vaccari, M. 2008, A\& A, 487, 837

\bibitem[Georganopoulos {et~al.}(2010)]{Georganopoulos2010}
Georganopoulos, M., Finke, J.~D. \& Reyes, L.~C. 2010, ApJ, 714, L157

\bibitem[Gilmore {et~al.}(2009)]{Gilmore2009}
Gilmore, R., Madau, P., Primack, J.~R., {et~al.} 2009, MNRAS, 399, 1694

\bibitem[Gould \& Schr\'{e}der(1966)]{Gould1966}
Donea, A.-C. \& Schr\'{e}der, R.~J. 1966, Phys. Rev. Lett., 16, 252

\bibitem[Hartman {et~al.}(1999)]{Hartman1999}
Hartman, R.~C., Bertsch, D.~L., Bloom, S.~D., {et~al.} 1999, ApJS, 123, 79

\bibitem[Hauser {et~al.}(2004)]{Hauser2004}
Hauser, M., M{\"o}llenhoff, C., P{\"u}hlhofer, G., {et~al.} 2004, Astronomische  Nachrichten, 325, 659

\bibitem[Hinton (2004)]{Hinton2004}
Hinton, J. 2004, New. Astron. Review, 48, 331

\bibitem[Homan {et~al.}(2001)]{Homan2001}
Homan, D.~C., Ojha, R., Wardle, J. F.~C., {et~al.} 2001, ApJ, 549, 840

\bibitem[Homan {et~al.}(2002)]{Homan2002}
Homan, D.~C., Wardle, J. F.~C., Cheung, C.~C., {et~al.} 2002, ApJ, 580, 742

\bibitem[Jorstad {et~al.}(2006)]{Jorstad2006}
Jorstad, S.~G., Marscher, A.~P., Aller, M.~F. \& Balonek, T.~J. 2006, ASPC,  360, 169 - ASP Conference Series: \emph{AGN Variability from X-Rays to Radio Waves}

\bibitem[Jorstad {et~al.}(2005)]{Jorstad2005}
Jorstad, S.~G., Marscher, A.~P., Lister, M.~L., {et~al.} 2005, AJ, 130, 1418

\bibitem[Kalberla {et~al.}(2005)]{Kalberla2005}
Kalberla, P. M.~W., Burton, W.~B., Hartmann, D., {et~al.} 2005, A\&A, 440, 775

\bibitem[Kataoka {et~al.}(2008)]{Kataoka2008}
Kataoka, J., Madejski, G., Sikora, M., {et~al.} 2008, ApJ, 672, 787

\bibitem[Larionov {et~al.}(2009)]{Larionov2009}
Larionov, V.~M., Villata, M., Raiteri, C.~M., {et~al.} 2009, ATel, 1990

\bibitem[Li \& Ma(1983)]{LiMa1983}
Li, T.-P. \& Ma, Y.-Q. 1983, ApJ, 272, 317

\bibitem[Liu \& Bai(2006)]{LiuBai2006}
Liu, H.~T. \& Bai, J.~M. 2006, ApJ, 653, 1089

\bibitem[Malkan \& Moore(1986)]{MM1986}
Malkan, M.~A. \& Moore, R.~L. 1986, ApJ, 300, 216

\bibitem[Mattox {et~al.}(1996)]{Mattox1996}
Mattox, J.~R., Bertsh, D.~L., Chiang, J., {et~al.} 1996, ApJ, 461, 396

\bibitem[Marscher {et~al.}(2010)]{Marscher2010}
Marscher, A.~P., Jorstad, S.~G., Larinov, V.~M., {et~al.} 2010, ApJ, 710, L126

\bibitem[Mazin \& Raue (2007)]{Mazin2007}
Mazin, D. \& Raue, M. 2007, A\&A, 471, 439

\bibitem[Mead {et~al.}(1990)]{Mead1990}
Mead, A. R.~G., Ballard, K.~R., Brand, P. W. J.~L., {et~al.} 1990, A\&AS, 83,  183

\bibitem[Meyer {et~al.}(2012)]{Meyer2012}
Meyer, M., Raue, M., Mazin, D., {et~al.} 2012, A\&A, 542, A59

\bibitem[Nikishov(1962)]{Nikishov1962}
Nikishov, A.~I. 1962, Sov. Phys. J. Exp. Theor. Phys., 14, 393

\bibitem[{Nolan {et~al.}(2012)}]{2012Nolan2FGL}
Nolan, P.~A., Abdo, A.~A., Ackermann, M., {et~al.} (Fermi Collaboration) 2012, ApJS, 199, 31N

\bibitem[Ohm {et~al.}(2009)]{Ohm2009}
Ohm, S., van Eldik, C. \& Egberts, K. 2009, Astroparticle Physics, 31, 383

\bibitem[Piron {et~al.}(2001)]{Piron2001}
Piron, F., Djannati-Ata\"{i}, A., Punch, M., {et~al.} 2001, A\& A, 374, 895

\bibitem[Poutanen \& Stern (2010)]{PoutanenStern2010}
Poutanen, J., Stern, B. 2010, ApJ, 717, L118

\bibitem[Pucella {et~al.}(2009)]{Pucella2009}
Pucella, G., D'Ammando, F., Tavani, M., {et~al.} 2009, ATel, 1968

\bibitem[Salamon \& Stecker (1998)]{SalmonStecker1998}
Salamon, M. H. \& Stecker, F. W. 1998, ApJ, 493, 547

\bibitem[Stockman {et~al.}(1984)]{Stockman1984}
Stockman, H.~S., Moore, R.~L. \& Angel, J. R.~P. 1984, ApJ, 279, 485

\bibitem[Tadhunter {et al.}(1993)]{Tadhunter1993}
Tadhunter, C.~N., Morganti, R., Di~Serego-Alighieri, S. et al. 1993, MNRAS, 263, 999

\bibitem[Tavecchio {et al.}(2011)]{Tavecchio2011}
Tavecchio, F., Becerra-Gonzalez, J., Ghisellini, G. et al. 2011, A\&A, 534, A86

\bibitem[Tanaka {et al.} (2011)]{Tanaka2011}
Tanaka, Y.~T., Stawarz, \L., Thompson, D.~J. et al. 2011, ApJ, 733, 19

\bibitem[Vercellone {et~al.}(2009)]{Vercellone2009}
Vercellone, S., D'Ammando, F., Pucella, G., Tavani, M., {et~al.} 2009, ATel,  1976

\bibitem[Villata {et~al.}(2009)]{Villata2009}
Villata, M., Raiteri, C.~M., Larionov, V.~M., {et~al.} 2009, ATel, 1988

\bibitem[Wagner {et al.} (1995)]{WagnerS1995}
Wagner, S.~J., Camenzind, M., Dreissigacker, O., {et~al.} 1995, A\&A, 298, 688

\bibitem[Wagner \& Witzel(1995)]{WagnerSJ1995}
Wagner, S.~J. \& Witzel, A. 1995, Annu. Rev. Astron. Astrophys., 33, 163

\end{thebibliography}
\end{document}